\documentclass[aps,prl,twocolumn,showpacs]{revtex4}
\usepackage{epsf}
\usepackage{dcolumn}
\usepackage{amsmath}
\usepackage[dvips,xdvi]{graphicx,color}

\makeatletter
\def\@cite#1#2{$^{#1\if@tempswa , #2\fi}$\ }
\makeatother

\newif\iffigure
\figuretrue
\newif\iftable
\tabletrue

\newcounter{fgr}
\newcommand{\be}{\begin{equation}}
\newcommand{\ee}{\end{equation}}
\def\bal{\begin{align}}
\def\eal{\end{align}}
\newcommand{\ba}{\begin{eqnarray}}
\newcommand{\ea}{\end{eqnarray}}

\def\av#1{{\left\langle#1\right\rangle}}
\def\pt#1#2{{\partial#1\over\partial#2}}

\def\dx#1#2{{d#1\over d#2}}
\def\rf#1{(\ref{#1})}

\def\Rl{R_{\lambda}}
\def\eb{\bar{\epsilon}}

\def\A{{\mbox{\boldmath$A$}}}
\def\e{{\mbox{\boldmath$e$}}}
\def\H{{\mbox{\boldmath$H$}}}
\def\r{{\mbox{\boldmath$r$}}}
\def\u{{\mbox{\boldmath$u$}}}
\def\V{{\mbox{\boldmath$V$}}}
\def\w{{\mbox{\boldmath$w$}}}
\def\x{{\mbox{\boldmath$x$}}}
\def\X{{\mbox{\boldmath$X$}}}
\def\y{{\mbox{\boldmath$y$}}}

\def\Gam{{\mit{\Gamma}}}
\def\ob{{\mbox{\boldmath$\omega$}}}

\begin{document}

\title{\vbox to 0pt {\vskip -1cm \rlap{\hbox to \textwidth {\rm{\small}
\hfill 
}}}Role of pressure in turbulence
}

\title[Role of pressure in turbulence]{Role of pressure in turbulence
}

\author{\bf Toshiyuki Gotoh}
\email{gotoh@system.nitech.ac.jp}
\affiliation{Department of Systems Engineering, Nagoya Institute of Technology, \\ 
Showa-ku, Nagoya 466-8555, Japan}

\author{\bf  and Tohru Nakano}
\affiliation{Department of Physics, Chuo University, 
Kasuga, Bunkyo-ku,  Tokyo, 112-8551, Japan \\
}


\begin{abstract}\noindent

\vspace*{0.5cm}
There is very limited knowledge of the kinematical relations for the  
velocity structure functions higher than three. Instead, 
the dynamical equations for the structure functions of the 
velocity increment are obtained from the Navier Stokes equation under 
the assumption of the local homogeneity and isotropy. 
These equations contain the correlation 
between the velocity and pressure 
gradient increments, which is very difficult to know theoretically and 
experimentally. We have examined these dynamical relations by using 
direct numerical simulation data at very high resolution 
at large Reynolds numbers, 
and found that the contribution of the pressure term is important 
to the dynamics of the longitudinal velocity with large amplitudes. 
The pressure term is examined from the view point of 
the conditional average and the role of the pressure term in the 
turbulence dynamics is discussed. 

\end{abstract}
\maketitle


\section{Introduction}

In Kolmogorov's theory (1941, hereafter K41), 
the moments of velocity increments in the isotropic turbulence scale as 
$\av{U^p}\propto r^{\zeta_p}$ with $\zeta_p=p/3$,  
where $U$ is the longitudinal increment defined as 
$U=(\u(\x+\r)-\u(\x))\cdot\r/r$\cite{Kol41,My72}. 
However, it is now widely accepted from the experimental and 
direct numerical simulation (DNS) data  
that at high Reynolds numbers the intermittency builds up 
with decrease of scales and scaling exponent $\zeta_p$ does not 
obey K41 and is a non-decreasing function of $p$. 

Arguments have recently been raised 
as to whether or not the scaling exponent of the longitudinal structure function 
is equal to that of the transverse one at the order higher than three 
\cite{Lvovetal97,Heetal98,Dhruvaetal97,Wateretal99,Heetal99,Nelkin99,Sw01}.  
The structure function (moment) of the velocity increment is defined by 
\be 
   S_{m,n}(r)=\av{U^mV^n},         \label{1}
\ee
where $V$ is the transverse velocity increment. 
In the inertial range at very high Reynolds numbers, $S_{m,n}(r)$ is assumed 
to obey power law $S_{m,n}(r)\propto r^{\zeta_{m,n}}$.  
In the homogeneous isotropic turbulence, the incompressible 
condition yields the well known relations at the second and third order 
\begin{align}
 S_{0,2}(r) & =S_{2,0}(r)+{r\over 2}{d\over dr}S_{2,0},\label{2}\\
 S_{1,2}(r) & ={1\over 6}S_{3,0}(r)+{r\over 6}{d\over dr}S_{3,0}(r), \label{3}
\end{align}
from which  it follows $\zeta_{2,0}=\zeta_{0,2}$ and 
$\zeta_{3,0}=\zeta_{1,2}$. Kolmogorov's $4/5$ and $4/3$ laws are 
derived from the Navier-Stokes equation as 
\be
     S_{3,0}(r)=-{4\over 5}\eb r,    \quad 
     S_{3,0}(r)+2S_{1,2}(r)=-{4\over 3}\eb r,             \label{4}
\ee
so that $\zeta_{3,0}=\zeta_{1,2}=1$, 
where $\eb$ is the average rate of energy dissipation rate per unit mass.  
However, for the order higher than three, no equations are known 
that tie directly the longitudinal structure functions to 
the transverse or mixed structure functions. 
Instead, relations among various type of structure functions are 
derived from the Navier-Stokes equation which contains the pressure and 
viscous terms\cite{Fuka00,Yakhot01,Hill01} such as 
\begin{equation}
\av{(\delta_r(\nabla_{\parallel}p))U^pV^q}, \quad 
    \av{(\delta_r(\nabla^2U))U^pV^q}  \label{4b} 
\end{equation}
where 
\begin{eqnarray}
&& \delta_r(\nabla_{\parallel}p)
    =\left(\pt{p(\x_1)}{\x_1}-\pt{p(\x_2)}{\x_2}\right)\cdot{\r\over r}, 
                                                          \label{4c}\\
&& \delta_r(\nabla^2U)
    =\left(\nabla^2_{\x_1}\u(\x_1)
                     -\nabla^2_{\x_2}\u(\x_2)\right)\cdot{\r\over r}. 
                                                          \label{4d}
\end{eqnarray}

Gotoh considered the pressure term from the view point of 
the conditional average of the longitudinal increments of the pressure 
gradients $\av{\delta \nabla_{\parallel}p|U,r}$ for given value of $U$ 
(see later sections for more detail)\cite{Gotoh00}. The conditional average 
is expressed in terms of the quadratic function of $U$ and the coefficients 
are fixed by some statistical constraints. The qualitative feature of the 
conditional average was compared with DNS data and the results was 
found to be satisfactory. 

Yakhot studied the equations of the structure functions 
by using the generating functions which had been introduced by 
Polyakov\cite{Poly95,Yakhot01}. 
The same equations were obtained also by Hill by directly dealing with 
the velocity increments instead of the generating function\cite{Hill01}. 
In the equations, the terms arising from the convective term of 
the Navier Stokes equation are expressed in terms of $S_{p,q}$ alone, 
but the equations again contain contributions from the pressure and 
viscous terms. Yakhot argued that in the equations of $S_{2n,0}$ 
the contribution from the viscous term vanishes and
the pressure term can be neglected, and concluded 
$\zeta_{2n,0}=\zeta_{2n-2,2}$. 
For the equations of $S_{p,2m}$, the pressure term is modelled as 
the conditional average of the pressure gradient increments 
for given values of $U$ and $V$,  
$\av{\delta \nabla_{\perp}p|U,V,r}$, 
which is assumed to be given by the expansion 
in velocity increments and more general than Gotoh's case. 
He retained terms up to the second order, and argued a resemblance 
to ``mean-field approximation" in the critical phenomena\cite{Yakhot01}. 
This model for the pressure term enabled him to close the equation 
for the probability density function (PDF) of $V$ in the inverse cascading 
range in the two dimensional turbulence where the dissipation 
term is irrelevant.  The PDF is found to be Gaussian 
and the normal scaling for $S_{0,q}$ was obtained.  
The same closure for the pressure term and a model for the dissipation term 
yield a closed equation for the PDF of $V$ in three dimensional turbulence, 
from which the peak value and the asymptotic tail of the PDF are derived, 
and the scaling exponent for $S_{0,q}$ is computed. 

These facts suggest that when the dissipation contributions to 
the equation of the velocity structure functions are irrelevant 
the pressure-velocity correlation is 
a key to understand the anomalous scaling exponents. 
The average of the pressure term conditioned upon the velocity 
increment is a kind of statistical projection of the pressure force 
on the dynamics of the velocity increment and study of its functional 
form would yield important knowledge of the scaling of the structure 
functions\cite{Yakhot01,Hill01}. 

Kurien and Sureenivasan studied experimentally the equations of the 
above structure functions by using the turbulence data measured in 
the atmospheric boundary layer at the Talyor microscale Reynolds 
number $R_{\lambda}\approx 10700$. The width of the inertial range 
is about one decade. Since the direct measurement of the contributions 
of the pressure and viscous terms are difficult, they considered first 
the equations of the structure functions without the pressure and 
viscous terms, and then examined the balance among the terms 
in the equations. They concluded that the pressure effects 
are less important for $S_{2n,0}$ at low order\cite{Kurien01}. 
Further, in order to assess the contributions of the pressure term, 
they applied Yakhot's theory to the pressure term. With this model, 
the correlation between the velocity increments and the pressure 
gradient increment is expressed in terms of the velocity structure 
functions alone, meaning a closure. 
They concluded that the pressure and viscous terms are small 
compared to the other terms consisting of $S_{p,q}$. 
Also the scaling exponents of $S_{0,q}$, the peak values of the PDF and 
the asymptotic tail of the PDF of $V$ were examined and compared to 
Yakhot's prediction. An agreement was observed for these quantities, 
but insufficient for the pressure contributions. 
Since there is small but finite degree 
of anisotropy at second and third order structure functions, and 
the data in the experiment suffers from noise, the conclusion 
is very suggestive but inconclusive. Moreover the pressure and viscous terms 
are examined indirectly through the assumed model to the pressure term. 
Certainly we are left with lack of precise knowledge for the equation and 
further study of the equation of $S_{p,q}$ is highly needed. 
Especillay the direct examination of the pressure and viscous terms is 
required. 

The above story motivates us to examine the dynamical relations 
for $S_{m,n}(r)$ in terms of DNS. DNS can provide precise data of 
each term in the equations so that the contributions of the pressure and 
viscous terms are directly examined. 
When compared to the atmospheric boundary layer data, $\Rl=460$ of the DNS 
is low, but degree of isotropy is so well that about one decade of the 
inertial range width is achieved\cite{Gotohetal02}.  
In the following, the analysis of the DNS data shows that the pressure term 
is important to the dynamical equations of the longitudinal correlation 
functions. Conditional average of the pressure gradient increments is also 
analysed and its role in the dynamics is discussed. 

\section{Dynamical relations}

The scalewise energy budget is well described by 
the K\'arman-Howarth-Kolmogorov equation 
\be
  {4\over 5}\eb r= -S_{3,0}+ 6\nu {dS_{2,0}\over dr}
                                      +Z_{force}(r),  \label{5} 
\ee
where $Z_{force}(r)$ is an energy input 
by an external force. 
Note that this equation does not contain the pressure term. 
When the external force has a support compact spectrum at low
wavenumbers, both the viscous and external force terms can be 
neglected so that we have Kolmogorov's $4/5$ law.  

The dynamical relations at higher order in the inertial range are 
derived from the Navier-Stokes equation. This can be efficiently 
done by introducing the generating function 
$Z=\av{\exp(\lambda_{\alpha}w_{\alpha})}$ 
where $\w=\u(\x_1)-\u(\x_2)$ is the velocity difference
\cite{Poly95,Yakhot01}. The resulting equation is 
\begin{eqnarray}
&& \hspace*{-0.8cm} 
\dx{S_{p,q}}{r}+{d-1+q\over r}S_{p,q}             
       -{(p-1)(q-1+d)\over r(q+1)}S_{p-2,q+2}        \nonumber\\
&& \hskip 0.5cm 
   = -(p-1)\av{(\delta \nabla_{\parallel}p)U^{p-2}V^q}
\nonumber\\
 && \hskip 1.5cm             
        -q\av{(\delta \nabla_{\perp}p)U^{p-1}V^{q-1}}-D_{p,q},  
                                                   \label{11} 
\end{eqnarray}
where  $d$ is the dimensionality. $D_{p,q}$ denotes the contributions 
from the viscous terms 
\begin{eqnarray}                                              
&& D_{p,q}=\nu\Bigl((p-1)(p-2)\av{(\nabla_{\X}U)^2U^{p-3}V^q}
\nonumber\\
 &&   \hskip 0.5cm
        +2q(p-1)\av{(\nabla_{\X}U)\cdot(\nabla_{\X}V)U^{p-2}V^{q-1}}  
                        \nonumber\\
&& \hskip 1.0cm     +q(q-1)\av{(\nabla_{\X}V)^2U^{p-1}V^{q-2}}\Bigr),  
                                                   \label{11b}                                                   
\end{eqnarray}
where $\X=(\x_1+\x_2)/2$, 
$\nabla_{\X}U=\partial u_l(\x_1)/\partial \x_1-\partial u_l(\x_2)/\partial \x_2$, 
$\nabla_{\X}V=\partial u_t(\x_1)/\partial \x_1-\partial u_t(\x_2)/\partial \x_2$, 
and $u_l$ and $u_t$  are the longitudinal and transverse 
components of the velocity vector, respectively. The external force term is also   
neglected\cite{Yakhot01,Hill01}, and the pressure terms appear in Eq.\rf{11}. 

In the inertial range, 
Yakhot, Hill and Boratav, Hill, and Kurien and Sreenivasan have discussed 
Eq. \rf{11} for some set of $(p,q)$ without pressure and viscous contributions
\cite{Yakhot01,HillBora01,Hill01,Kurien01}. 
Following their discussion and for latter convenience 
we write a first few series of Eqs.\rf{11} without the pressure and 
viscous terms as 
\begin{align}
  & \dx{S_{4,0}}{r}+{2\over r}S_{4,0}={6\over r}S_{2,2}, \label{12}\\
  & \dx{S_{6,0}}{r}+{2\over r}S_{6,0}={10\over r}S_{4,2}, \label{13}\\
  & \dx{S_{8,0}}{r}+{2\over r}S_{8,0}={14\over r}S_{6,2},  \label{14}
\end{align}
for the longitudinal structure functions and 
\begin{align}
  & \dx{S_{2,2}}{r}+{4\over r}S_{2,2}={4\over 3r}S_{0,4}, \label{15}\\
  & \dx{S_{2,4}}{r}+{6\over r}S_{2,4}={6\over 5r}S_{0,6}, \label{16}\\
  & \dx{S_{4,2}}{r}+{4\over r}S_{4,2}={4\over r}S_{2,4},  \label{17}
\end{align}
for the mixed ones\cite{Hill01}. 
If the pressure and viscous terms can be neglected in the inertial range, 
we obtain the relations between the scaling exponents of the longitudinal, 
mixed, and transverse structure functions as $\zeta_{2n,0}=\zeta_{2n-2,2}$, 
$\zeta_{2,2n}=\zeta_{0,2n+2}$, and so on. 
In the following we examine the above relations 
quantitatively in terms of the DNS data.

\section{Examination with DNS data}

The turbulence in a periodic cubic box with size of $2\pi$  
is simulated by using the Fourier spectral method for space, 
and  Runge Kutta Gill method for time advancing.  
The Gaussian, white-in-time random solenoidal force is applied to 
the first few shells in the wavevector space to maintain  
statistically steady state. 
The total number of grid points is $N=1024^3$ and $k_{max}\eta=0.96$, 
where $\eta=(\nu^3/\eb)^{1/4}$ is the Kolmogorov scale. 
The value $0.96$ is slightly smaller than unity, but this does not affect 
the statistics of the inertial range. The ensemble average is taken over time 
average with the period of about four large eddy turnover time. 
Detail of the simulation is found in Gotoh et al.\cite{Gf01,Gotohetal02}. 
\begin{figure}[t]
\refstepcounter{fgr}
\label{gotoh_fig\thefgr}
{\hspace*{-0.8cm}
\includegraphics[width=9.5cm]{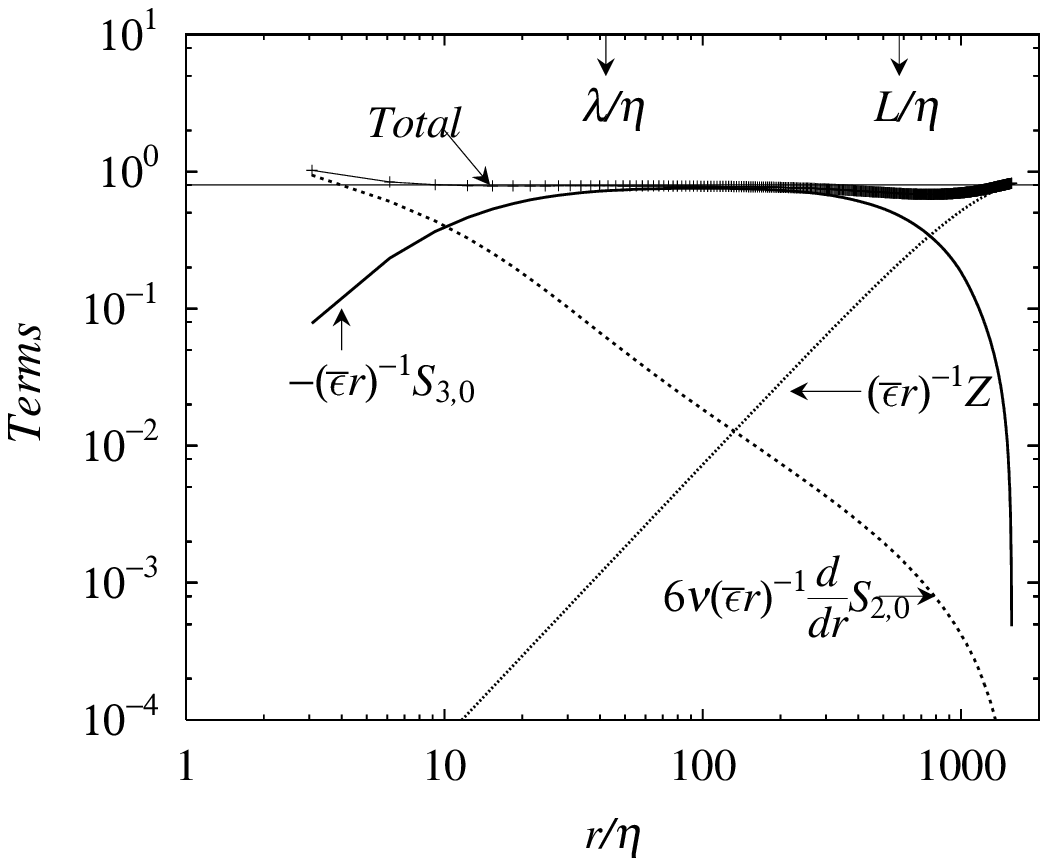}}
\vspace*{-5pt}
\centerline{\small Fig.\thefgr \ \ $4/5$-law at $R_{\lambda}=460$\cite{Gotohetal02}. }
\end{figure}%

Figure \ref{gotoh_fig1} shows each term of the 
K\'arman-Howarth-Kolmogorov equation \rf{5} divided by $\eb r$. 
When $r$ decreases, the forcing term decreases and 
$-S_{3,0}(r)$ quickly rises to the $4/5$ 
level and decays in the dissipation range. The energy budget 
follows Kolmogorov's 1941 theory and the inertial range 
certainly exists. 

In Figs. \ref{gotoh_fig2}, \ref{gotoh_fig3} and 
\ref{gotoh_fig4}, 
both sides of Eqs. \rf{12}, \rf{13} and \rf{14} 
are plotted, respectively, and the ratio of the difference of 
both sides to the left hand side ($|l.h.s.-r.h.s|/|l.h.s.|$) 
is also shown in the figures. 
If the contribution from the pressure gradient is negligible, 
both curves must collapse at least over a part or whole of the 
inertial range. Figures show that there exist differences and 
the left hand side of Eq. \rf{12} in Fig.\ref{gotoh_fig2} is bigger 
by factor 1.2 than the right hand side of Eq. \rf{12}. 
Distance between the two curves is almost constant over the inertial 
range, but increases with the order $n$. 
Corresponding to this, the ratio of the difference between 
both sides to the left hand side is nearly constant over 
the inertial range, which suggests that the contribution from 
the pressure gradient term scales almost in the same way as $S_{2n,0}$. 

When the pressure term is included in the right hand side of  
Eqs. \rf{12}, \rf{13} and \rf{14}, the curve of the right hand side 
(dash dotted line with plus symbol) perfectly collapses with 
the curve of the left hand side,  
which means that the pressure gradient term has the same order of 
contributions as the mixed correlation term. 
We have examined the contributions from the dissipation 
term and found that they are very small compared to the other terms 
of Eq. \rf{11} with $p=2m$ and $q=0$, consistent with the theoretical 
prediction\cite{Yakhot01}. 
For the dynamical equations of the odd order moments such as $S_{2n+1,0}(r)$, 
all the terms would contribute equally but unfortunately we have not 
computed these terms. 
\vfill

\begin{figure}[t]

\refstepcounter{fgr}
\label{gotoh_fig\thefgr}
{\hspace*{-0.8cm}\includegraphics[width=9.5cm]{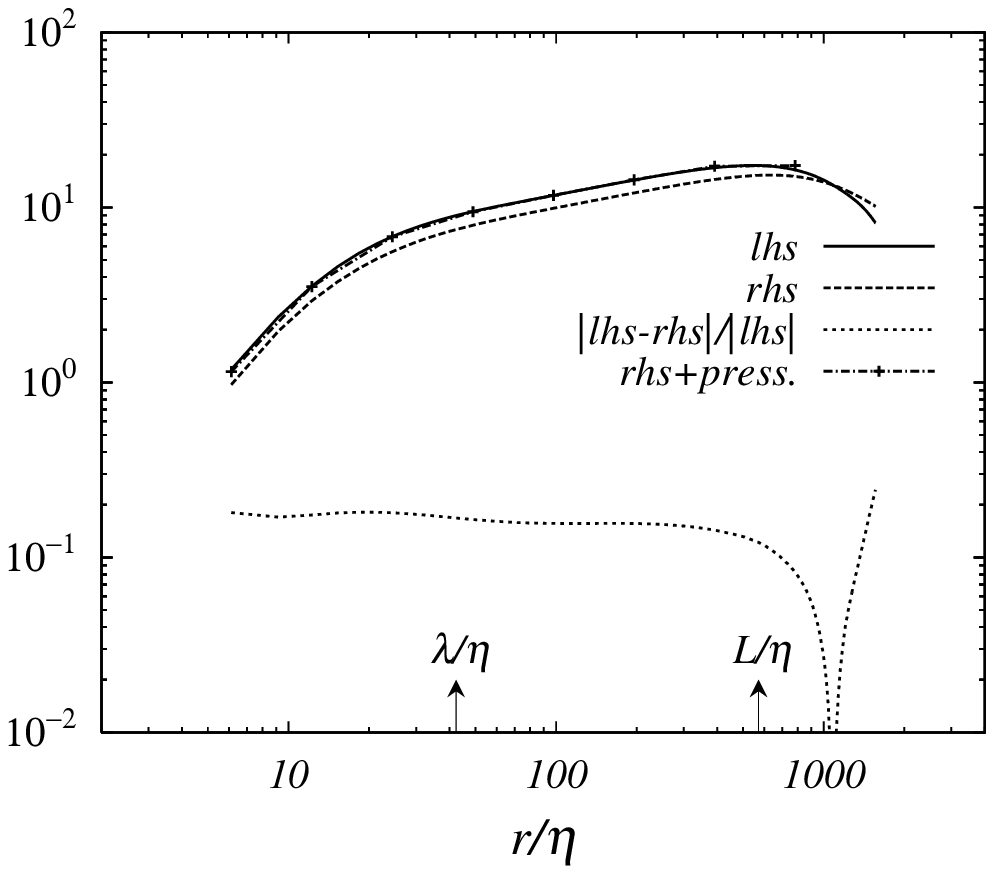}}
\vspace*{-5pt}
\centerline{\small Fig.\thefgr \ \ $\pt{S_{4,0}}{r}+{2\over r}S_{4,0}={6\over r}S_{2,2}
\ \ (-3\av{\delta_r p_xU^2})$. }
\end{figure}%

\begin{figure}[t]
\refstepcounter{fgr}
\label{gotoh_fig\thefgr}
{\hspace*{-0.8cm}\includegraphics[width=9.5cm]{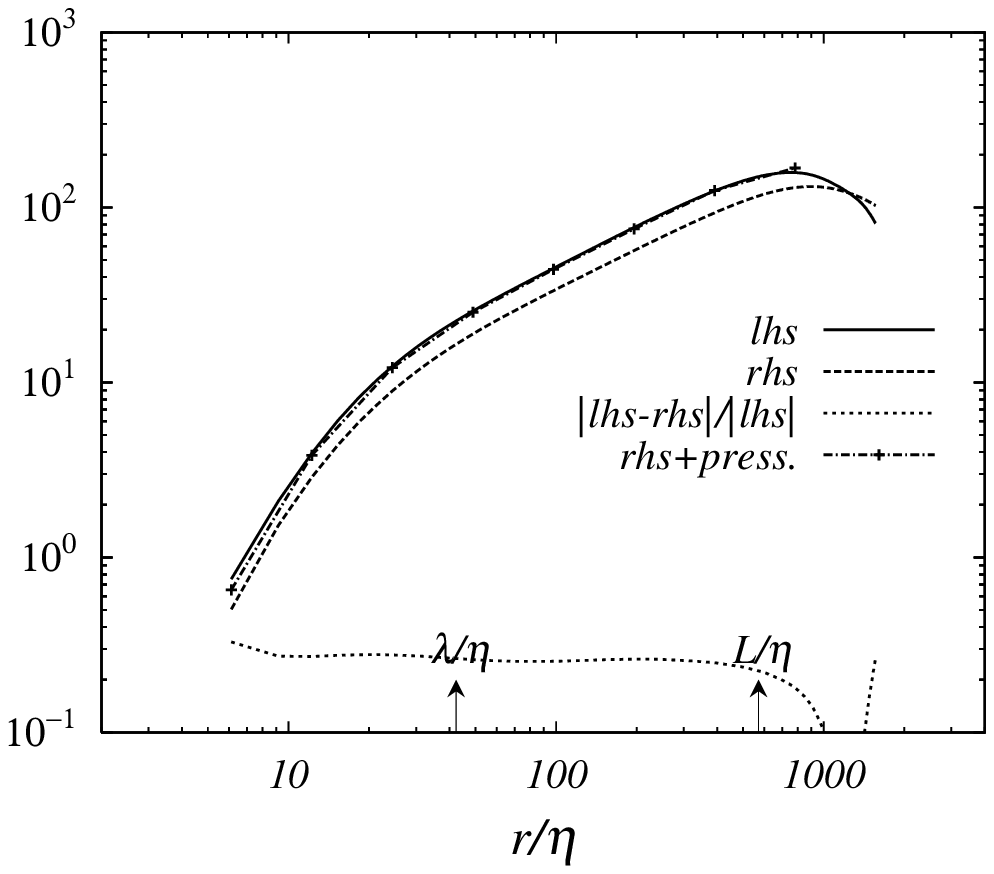}}
\vspace*{-5pt}
\centerline{\small Fig.\thefgr \ \ $\pt{S_{6,0}}{r}+{2\over r}S_{6,0}={10\over r}S_{4,2}
\ \ (-5\av{\delta_r p_xU^4})$.}
\end{figure}%

\begin{figure}[t]
\refstepcounter{fgr}
\label{gotoh_fig\thefgr}
{\hspace*{-0.8cm}\includegraphics[width=9.5cm]{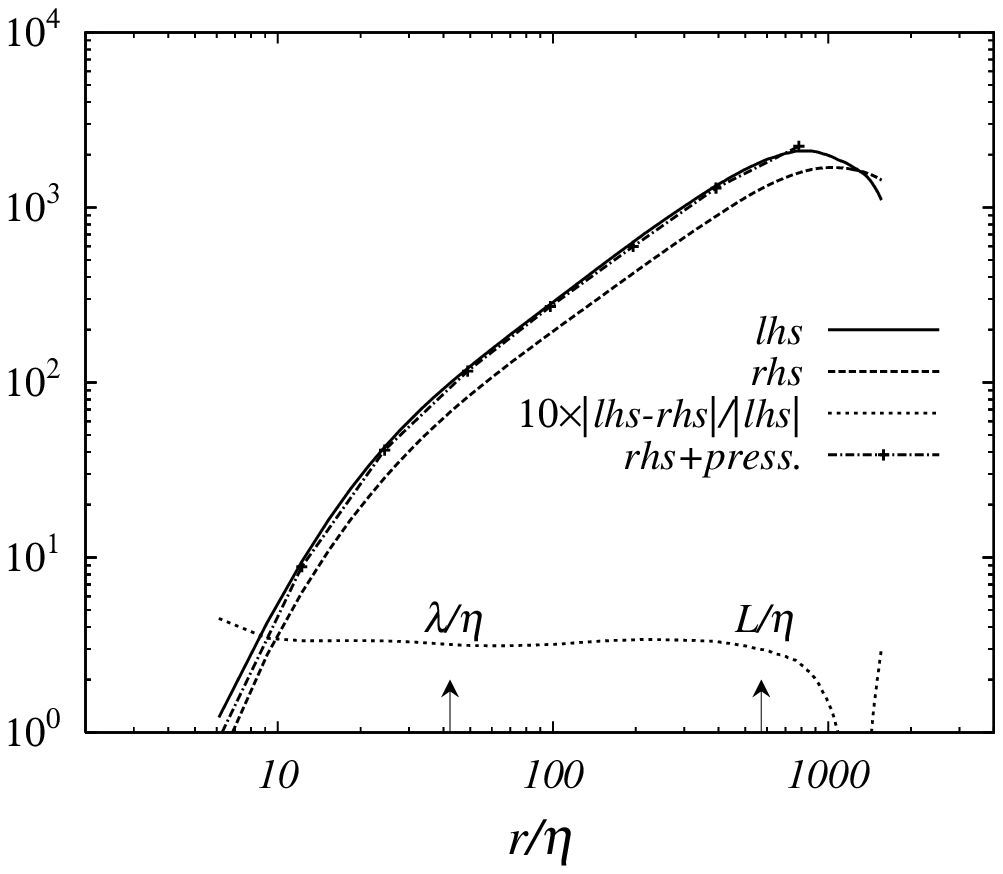}}
\vspace*{-5pt}
\centerline{\small Fig.\thefgr \ \ $\pt{S_{8,0}}{r}+{2\over r}S_{8,0}={14\over r}S_{6,2}
\ \ (-7\av{\delta_r p_xU^6})$. }
\end{figure}%
Figures \ref{gotoh_fig5}, \ref{gotoh_fig6} and 
\ref{gotoh_fig7} show the both sides of Eqs. \rf{15}, \rf{16} and \rf{17}. 
The curves for both sides of these equations are very close to each other 
for whole or part of the separation below the integral scale. 
Although it is not clear whether the difference between the curves remains 
to be small when the Reynolds number is increased, the degree of 
contribution of the pressure gradient to the dynamical equations in 
these cases is weaker than the case of the longitudinal case. 

{
\begin{figure}[t]
\refstepcounter{fgr}
\label{gotoh_fig\thefgr}
{\hspace*{-0.8cm}\includegraphics[width=9.5cm]{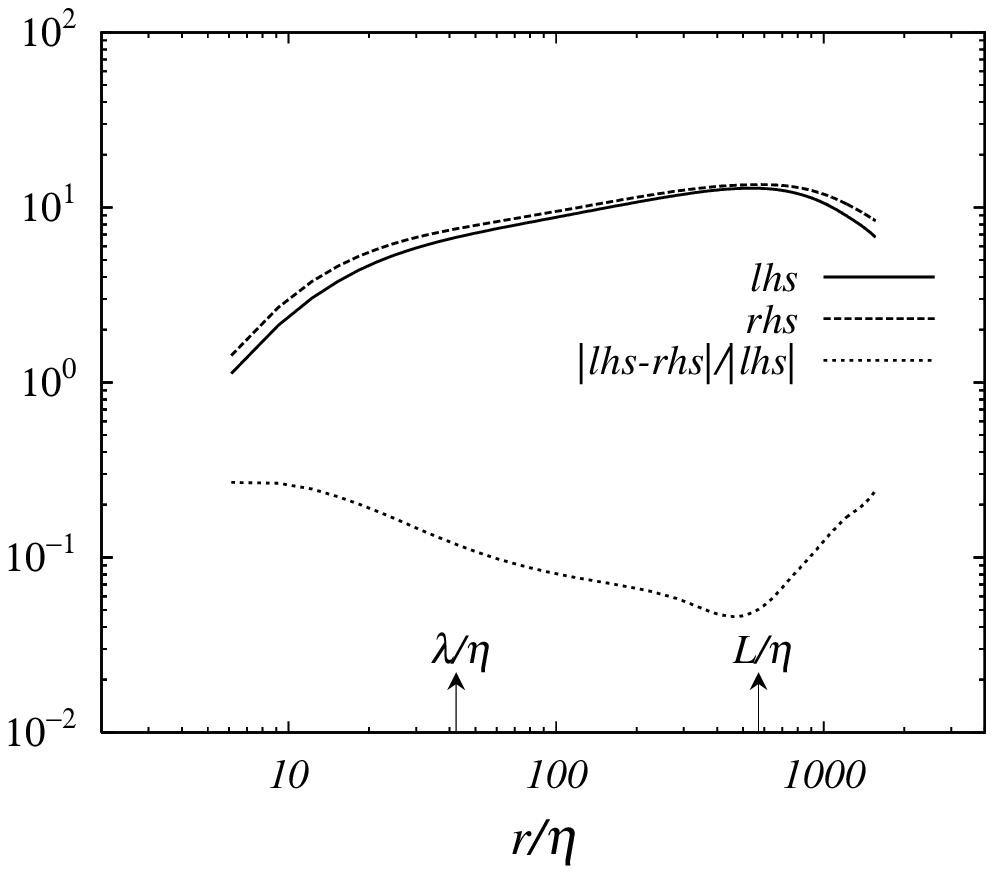}}
\vspace*{-5pt}
\centerline{\small Fig.\thefgr \ \ 
$ \pt{S_{2,2}}{r}+{4\over r}S_{2,2}\approx {4\over 3r}S_{0,4}$   
}
\end{figure}%

\begin{figure}[t]
\refstepcounter{fgr}
\label{gotoh_fig\thefgr}
{\hspace*{-0.8cm}\includegraphics[width=9.5cm]{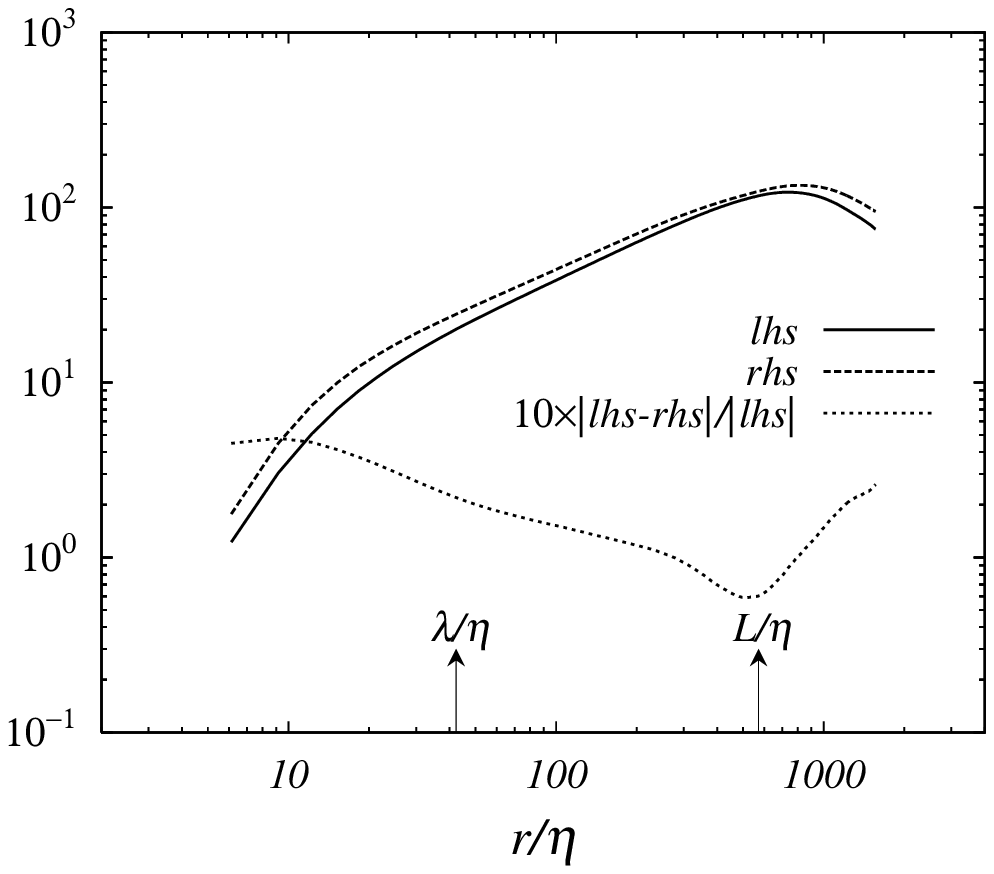}}
\vspace*{-5pt}
\centerline{\small Fig.\thefgr \ \ 
$\pt{S_{2,4}}{r}+{6\over r}S_{2,4}\approx {6\over 5r}S_{0,6}$ 
}
\end{figure}%

\begin{figure}[t]
\refstepcounter{fgr}
\label{gotoh_fig\thefgr}
{\hspace*{-0.8cm}\includegraphics[width=9.5cm]{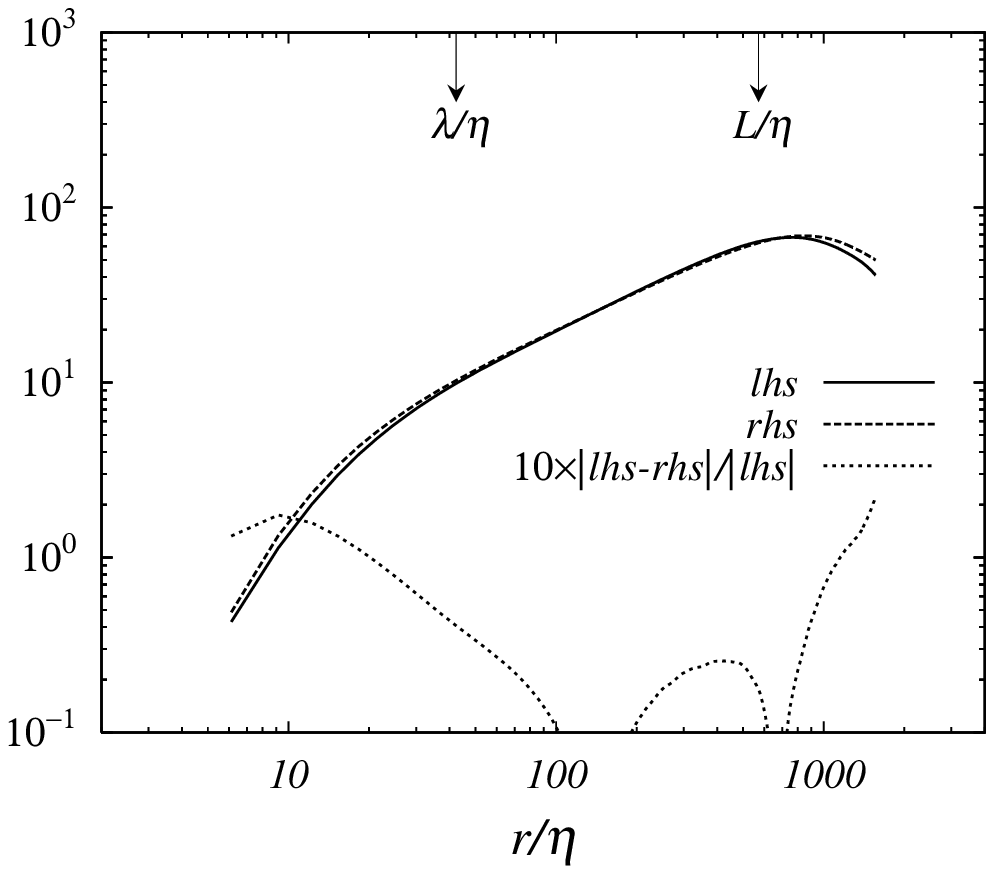}}
\vspace*{-5pt}
\centerline{\small Fig.\thefgr \ \ 
$\pt{S_{4,2}}{r}+{4\over r}S_{4,2}\approx {4\over r}S_{2,4}$}
\end{figure}%
}

These observations suggest that as far as the dynamics of the longitudinal 
velocity increments is concerned, the pressure term is very essential, 
while it is secondary for the dynamics of the transverse velocity increments.  
If the pressure term is absent, sharp shock is formed as 
in the Burgers turbulence, but in reality the pressure suppresses 
the formation of shock. This anti-shock action of the pressure would become 
stronger when $|U|$ with negative sign becomes larger. 
This implies that the larger the longitudinal 
velocity increments is, the more significant the 
contribution of the pressure term becomes (see more discussion in Sec.4). 
On the other hand, near the shock region, transverse velocity 
increments are closely tied to shearing motion as results  
of shock killer action of the pressure. 
Both positive and negative transverse velocity increment 
occur with equal probability so that 
the correlation between the transverse velocity increments and 
pressure gradient would have a good chance to be small. 

When the pressure term is included in the dynamical equation Eq.\rf{11}, 
it is not clear whether or not 
the scaling exponents $\zeta_{2n,0}$ and $\zeta_{2n-2,2}$ are equal. 
We need precise knowledge about the correlation between the increments 
of the pressure gradient and longitudinal velocity. 
In the following we examine the pressure term from the view point of 
the conditional average. 

\section{Conditional average of the pressure gradient increment}

The correlation function $\av{(\delta p_{,x})U^{2n-2}}$ can be written 
as follows 
\be
  \av{(\delta p_{,x})U^{2n-2}}
   =\int_{-\infty}^{\infty} \av{\delta p_{,x}|U,r}U^{2n-2}P(U,r)dU, \label{18}
\ee
where $\av{\delta p_{,x}|U,r}$ is the conditional average of the pressure 
gradient increment for a fixed value of $U$ and $P(U,r)$ is the PDF of  
$U$ with separation $r$. If $\av{\delta p_{,x}|U,r}$ is given as a 
polynomial function of $U$, then $\av{(\delta p_{,x})U^{2n-2}}$ is 
expressed in terms of the moments of $U$. For the moment we assume that 
$\av{\delta p_{,x}|U,r}$ is a quadratic function of $U$. Underlying 
motivation for this is that (1) the source term of the Poisson equation 
for the pressure is quadratic in the velocity gradient, i.e., 
when $u'=\gamma u$ we have $p'=\gamma^2 p$, and 
(2) for large $U$ the local Reynolds number is quite large so that 
the fluid motion would follow Bernoulli's theorem locally, 
implying that the pressure field is proportional to the square of 
the velocity. 
This is sketched as follows. 

Consider a vortex tube and a cylindrical 
surface surrounding the tube. On this surface, the velocity vector is 
nearly along the circumferential direction and the vorticity vector is 
approximately parallel to the axis of the vortex tube. 
Then the vector $\u\times\ob$ is parpendicular to the surface, 
the Bernoulli surface\cite{Bat67}. 
On this surface, the Bernoulli theorem is $p(\x_1)+\u^2(\x_1)/2=H_1$ 
where $\rho=1$ is assumed for simplicity and $H_1$ is a Bernoulli 
constant on the surface. Take the gradient, we have 
$-\nabla_1p_1=\u_1\cdot\nabla_1\u_1$. Then subtract from this  
equation the same expression evaluated at $\x_2$ on a different 
Bernoulli surface, and introduce the 
new cooridnate as $\X=(\x_1+\x_2)/2$ and $\r=\x_1-\x_2$, we obtain 
\ba
&& \hspace{-0.8cm} 
   -\delta \nabla p(\X,\r)
     =\V(\X,\r)\cdot\nabla_{\X}\delta\u(\X,\r) \nonumber\\
 && \hskip 1.5cm       
      +\delta\u(\X,\r)\cdot\nabla_{\r}\delta\u(\X,\r),    \label{18a}
\ea

\begin{figure}[t]
\refstepcounter{fgr}
\label{gotoh_fig\thefgr}
{\hspace*{-0.8cm}\includegraphics[width=9.5cm]{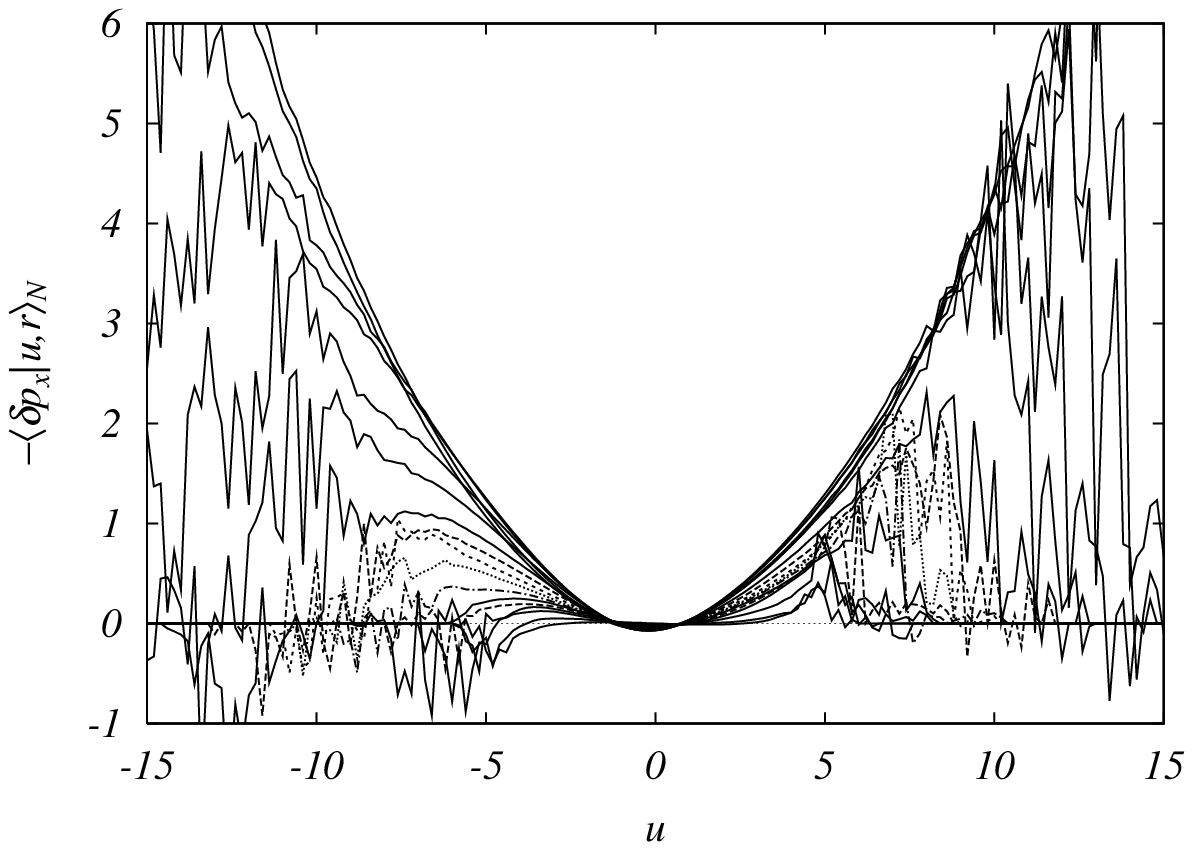}}
\vspace*{-5pt}
\centerline{\vbox{\small Fig.\thefgr \ \ 
$-\av{\delta p_x |u,r}_N$ for various $r/\eta$ 
($r_{min}/\eta=3$ and $r_{max}/\eta=1565.$). 
As the separation $r$ increases the curvature becomes smaller. }}
\end{figure}%

\begin{figure}[t]
\refstepcounter{fgr}
\label{gotoh_fig\thefgr}
{\hspace*{-0.8cm}\includegraphics[width=9.5cm]{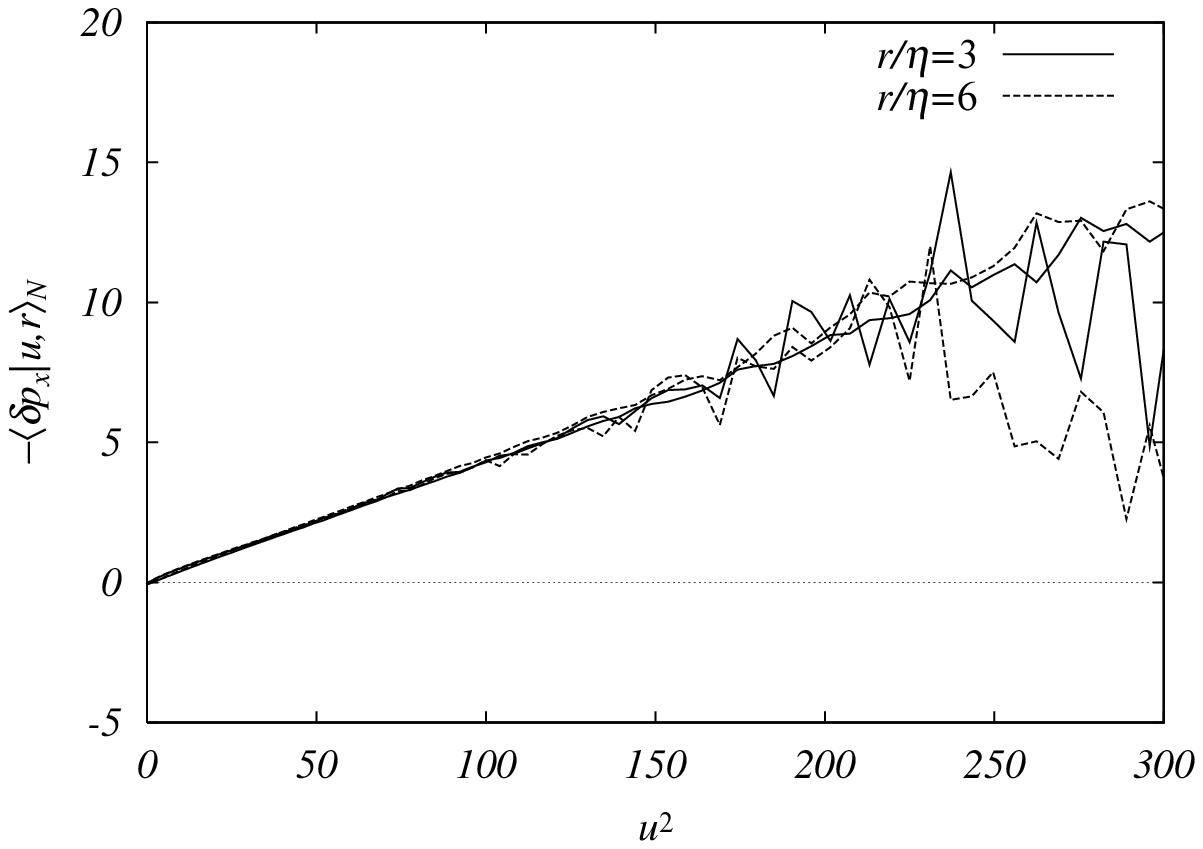}}
\vspace*{-5pt}
\centerline{\small Fig.\thefgr \ \ $-\av{\delta p_x |u,r}_N$ plotted against $u^2$ 
for $r/\eta=3$ and $6$.}
\end{figure}%

\noindent

\begin{figure}[t]
\refstepcounter{fgr}
\label{gotoh_fig\thefgr}
{\hspace*{-0.8cm}\includegraphics[width=9.3cm]{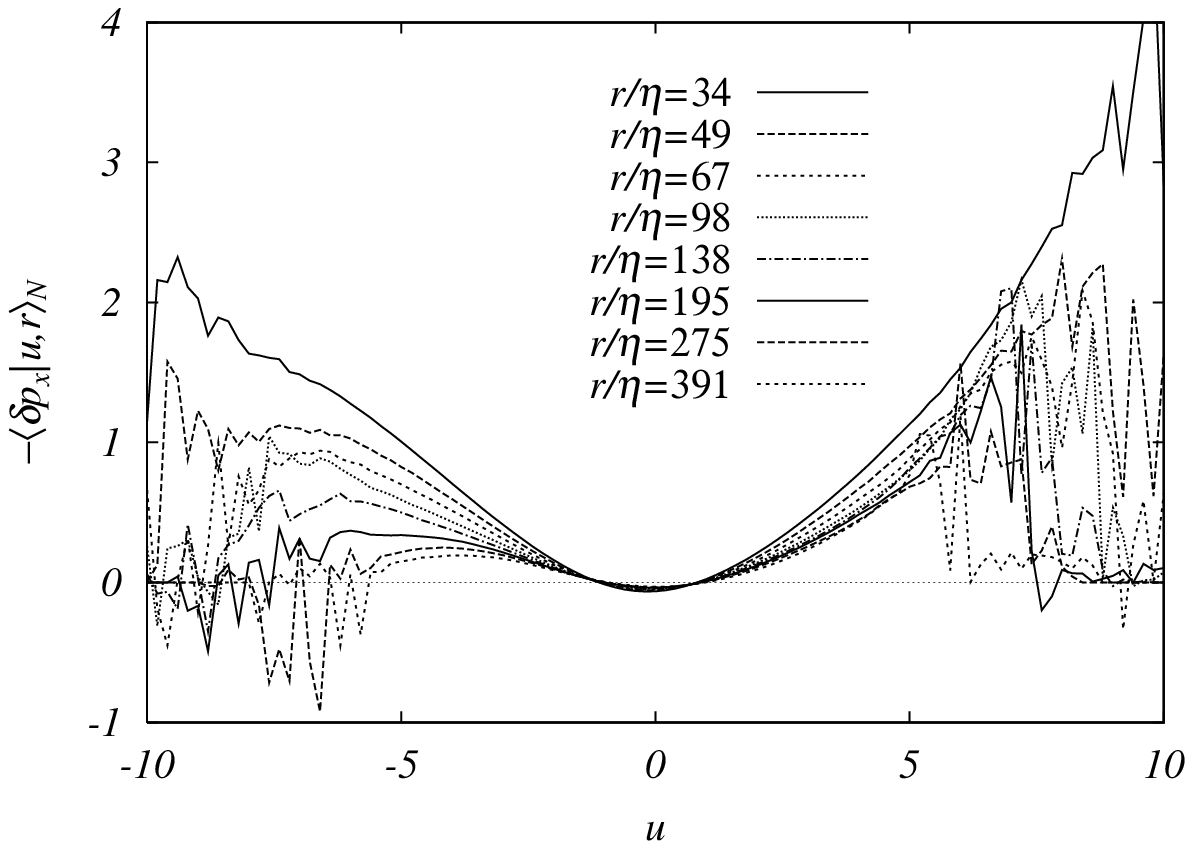}}
\vspace*{-5pt}
\centerline{\small Fig.\thefgr \ \ $-\av{\delta p_x |u,r}_N$ for $r/\eta$ 
in the inertial range. }
\end{figure}%

\begin{figure}[t]
\refstepcounter{fgr}
\label{gotoh_fig\thefgr}
{\hspace*{-0.8cm}\includegraphics[width=9.3cm]{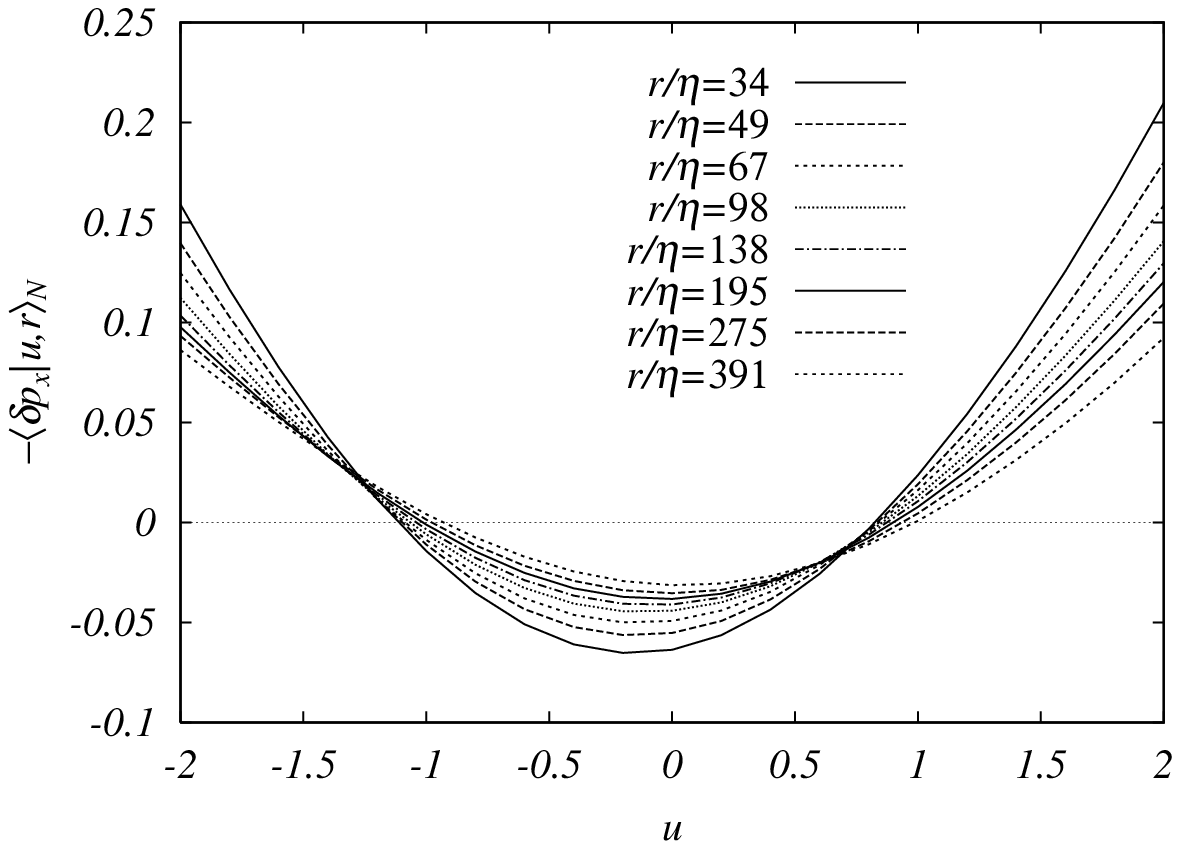}}
\vspace*{-5pt}
\centerline{\small Fig.\thefgr \ \ Close up of $-\av{\delta p_x |u,r}_N$ for $r/\eta$ 
in the inertial range.}
\end{figure}%

\noindent
where $\V(\x,\r)=(\u(\x_1)+\u(\x_2))/2$ which would be a slowly varing function 
of $\X$ and $\r$ compared to $\delta\u$. 
If we expect $\nabla_{\X}\delta\u(\X,\r)\propto \delta\u(\X,\r)$ and 
$\nabla_{\r}\delta\u(\X,\r)\propto \delta\u(\X,\r)$, 
Eq.\rf{18a} becomes quadratic in $\delta \u$. 
Moreover, since the Bernoulli constant differs on each Bernoulli 
surface, this fact would introduce an additional factor $\delta\H$,   
independent of the velocity difference, to Eq.\rf{18a}.  
The final expression would be 
$-\delta \nabla p=\A_0+\A_1\delta\u+\A_2\delta\u\delta\u$, where $\A_i$'s are 
tensor functions of $\X$ and $\r$. The average over $\X$ or statistical 
ensemble conditional upon the velocity increments would yield the 
quadratic function of $\delta\u$ for $-\av{\delta \nabla p|\delta\u,\r}$. 
The above argument is very rough and needs critical examination. 

The DNS data of the conditional average at  $R_{\lambda}=460$ is 
shown in Fig.\ref{gotoh_fig8} for the separation from the dissipation 
to the integral scale. In the plot, $u$ and $\av{\delta p_{,x}|u,r}_N$  
are normalized by their standard deviations, respectively. 
The curves are very close to the quadratic function of $u$. 
As the separation increases the curvature decreases, and the curves 
peel off the quadratic form at large amplitude of $u$. 
We infer that this is due to the fact that the number of large amplitude events 
becomes smaller as the amplitude $u$ increases. 

\begin{figure}[t]
\refstepcounter{fgr}
\label{gotoh_fig\thefgr}
{\hspace*{-0.8cm}\includegraphics[width=9.5cm]{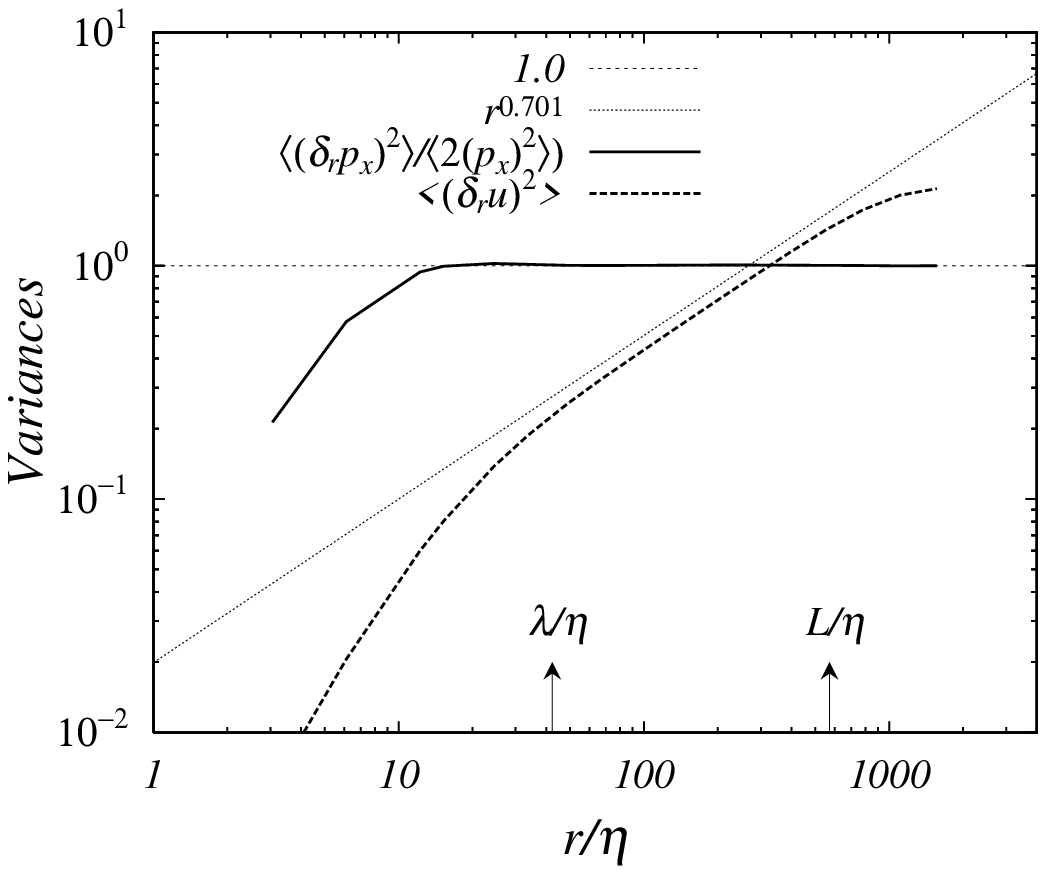}}
\vspace*{-5pt}
\centerline{\vbox{\small Fig.\thefgr \ \ Variation of the normalized variance of  
the pressure gradient increment 
${3\over 2}F_{\nabla p}^{-1}\eb^{-3/2}\nu^{1/2}\av{(\delta p_x)^2}$ 
with separation.}}
\end{figure}%

To see how close the curve is to the quadratic function of $u$, 
$\av{\delta p_{,x}|u,r}_N$ at $r/\eta=3$ and $6$ is plotted against $u^2$  
for both positive and negative $u$ in Fig. \ref{gotoh_fig9}. 
Curves are straight up to about $u^2\approx 200$, strongly suggesting 
that the curves are quadratic funtions of $u$. 
The curves for the inertial range separation are plotted in 
Fig.\ref{gotoh_fig10}. For small value of $u$, $\av{\delta p_{,x}|u,r}_N$  
is negative and positive at large $u$. Close inspection of 
Fig.\ref{gotoh_fig11} reveals that 
the vertex of the quadratic function locates at negative $u$ and moves 
towards the origin as the separation increases. 

It is very interesting to see how the quadratic function in $U$ for 
$\av{\delta p_{,x}|U,r}$ is fixed. The assumed quadratic form is  
\be 
 -\av{\delta p_{,x}|U,r} = a_2(r)U^2+a_1(r)U+a_0(r).   \label{19}
\ee
From the condition $\av{\delta p_{,x}}=0$, we have $a_0=-a_2S_{2,0}(r)$ 
where we used  $\av{U}=0$ and $\av{U^2}=S_{2,0}(r)$. Also the condition  
$\av{U\delta p_{,x}}=0$ by the homogeneity, 
isotropy and incompressibility yields $a_1(r)\av{U^2}=-\av{U^3}a_2(r)$\cite{Hill97,Boff02}. 
Combining these two relations 
we obtain $\av{\delta p_{,x}|u,r}_N$ in non-dimensional form as 
\ba
&&\hspace*{ -0.5cm } 
  -\av{\delta p_{,x}|u,r}_N
                                  \nonumber\\
&& \hskip 0.5cm
     =-\av{(\delta p_{,x})^2}^{-1/2}\av{\delta p_{,x}|U,r}  \nonumber\\
&& \hskip 0.5cm
    ={\hat{a}_2(r)\over \av{(\delta p_{,x})^2}^{1/2}}{S_{2,0}(r)\over r}
                        \left[u^2-s_3(r)u-1\right],  \label{20}
\ea
where $\hat{a}_2(r)=ra_2(r)>0$, and $s_3(r)=S_{3,0}(r)/[S_{2,0}(r)]^{3/2}$ 
is the skewness factor. 
Since $S_{3,0}(r)=-(4/5)\eb r$ in the inertial range, we obtain 
\be
s_3(r)=-{4\over 5}C_0^{-3/2}\left({r\over L}\right)^{1-{3\over 2}\zeta_{2,0}}, 
                                                       \label{21}
\ee
where $S_{2,0}(r)=C_0\eb^{2/3}r^{2/3}(r/L)^{\zeta_{2,0}-2/3}$ is used,  
and $L$ is a macro length scale,  
$C_0={27\over 55}\Gam(1/3)K\approx 1.32K$, 
and $K$ is the Kolmogorov constant. We expect that 
$\hat{a}_2(r)$ tends to a positive universal constant in 
the inertial range. The variance 
$\av{(\delta p_{,x})^2}=2(\av{p^2_{,x}}-\av{p_{,x}(\x+r\e_1)p_{,x}(\x)})$ 
quickly approaches the value $2\av{p^2_{,x}}$ when $r$ becomes larger than 
the dissipation scale as seen in Fig.\ref{gotoh_fig12}, 
so that it is well approximated as 
\be
  \av{(\delta p_{,x})^2}
      ={2\over 3}\eb^{3/2}\nu^{-1/2}F_{\nabla p}(R_{\lambda}),  \label{22}
\ee
where $F_{\nabla p}$ is a nondimensional constant defined by
$\av{(\nabla p)^2}=\eb^{3/2}\nu^{-1/2}F_{\nabla p}$.  
$F_{\nabla p}(R_{\lambda})$ is $R_{\lambda}$ dependent 
for low to moderate $R_{\lambda}$ and slowly tends to a constant\cite{Gf01,Vothetal02}. 
Using these quantities we write the conditional average as 
\ba
&&\hspace*{ -0.5cm }
  -\av{\delta p_{,x}|u,r}_N
                \nonumber\\
 && \hskip 0.2cm 
        =c_2(r)\left[\left(u+{|s_3(r)|\over 2}\right)^2
               -\left({|s_3(r)|^2\over 4}+1\right)\right]      \nonumber\\
 && \hskip 0.2cm 
     =\chi\left({r\over L}\right)^{\zeta_{2,0}-1}
         \left[\left(u+{|s_3(r)|\over 2}\right)^2
               -\left({|s_3(r)|^2\over 4}+1\right)\right],    \nonumber\\
    &&                                                    \label{23}
\ea

\begin{figure}[t]
\refstepcounter{fgr}
\label{gotoh_fig\thefgr}
\centerline{\includegraphics[width=9.cm]{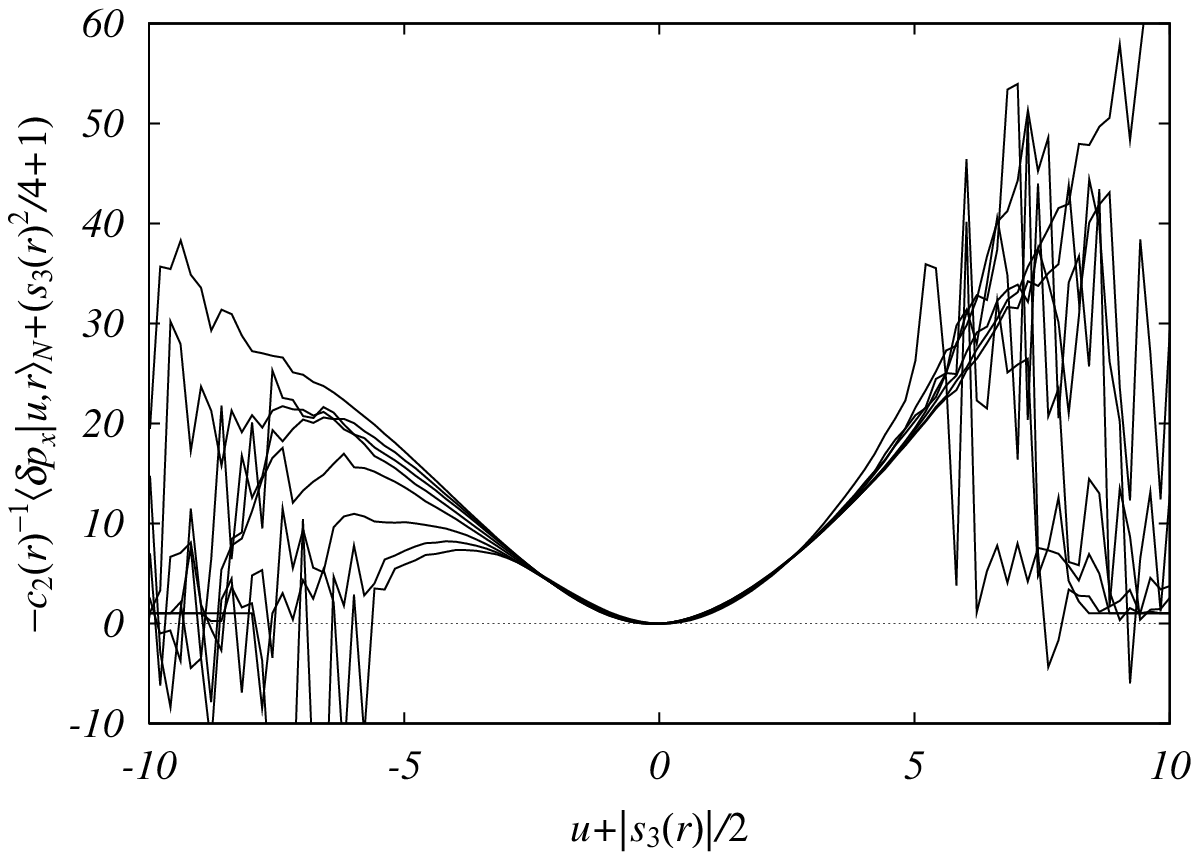}}
\vspace*{-5pt}
\centerline{\small Fig.\thefgr \ \ Scaled $-\av{\delta p_x |u,r}_N$.}
\end{figure}%

\noindent
in the inertial range, where 
\be
  \chi={AC_0\over \sqrt{(2/3)F_{\nabla p}(R_{\lambda})}}
          R_{\lambda}^{-1/2},    \label{24}
\ee
and  $A$ is a nondimensional constant. 
Figure \ref{gotoh_fig13} shows the scaled plot of 
$-c_2(r)^{-1}\av{\delta p_{,x}|u,r}_N+\left({|s_3(r)|^2\over 4}+1\right)$ against 
$u+{|s_3(r)|\over 2}$ for the inertial range separation. 
Collapse of curves for low to moderate amplitude is excellent. 

\begin{figure}[t]
\refstepcounter{fgr}
\label{gotoh_fig\thefgr}
{\hspace*{-0.8cm}\includegraphics[width=9.5cm]{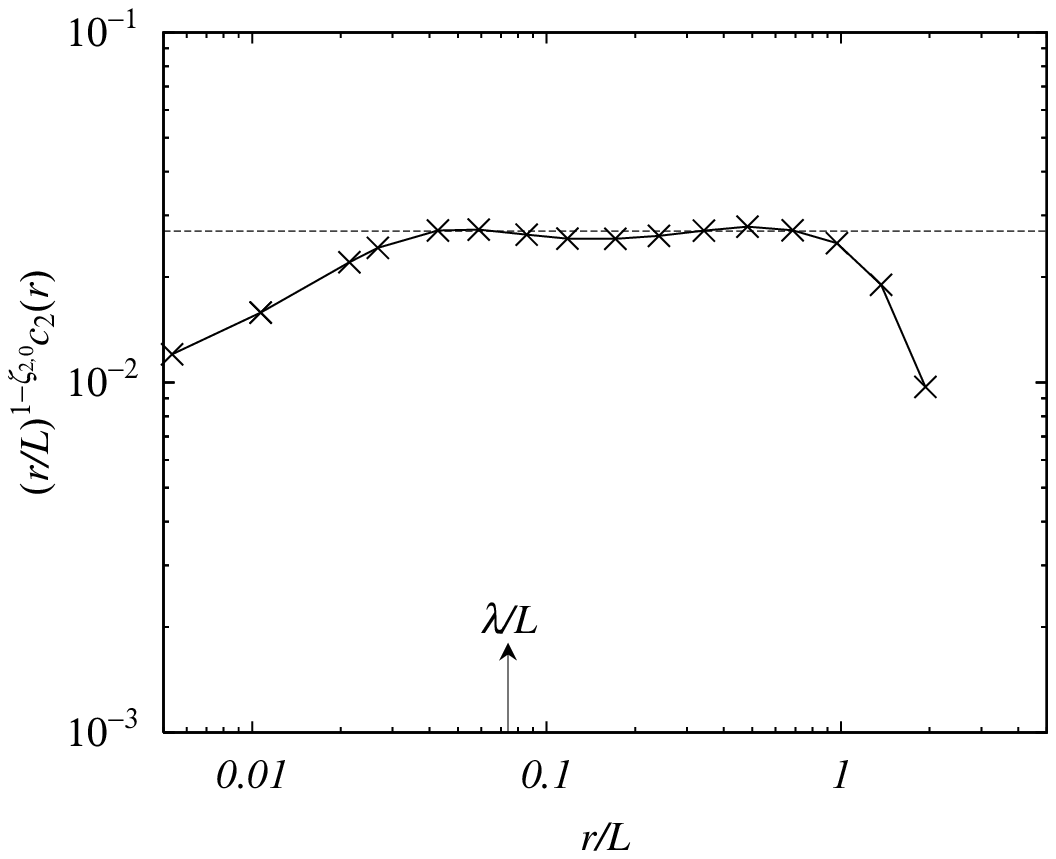}}
\vspace*{-5pt}
\centerline{\small Fig.\thefgr \ \ Compensated plot of the coefficients $c_2(r)$.}
\end{figure}%

We have computed the curvature of $-\av{\delta p_{,x}|u}_N$ at the origin 
and plotted in Fig.\ref{gotoh_fig14}. 
Since $c_2(r)$ is assumed to obey the power law $c_2(r)=\chi(r/L)^{-\alpha}$ 
in the inertial range, the curve is multiplied by $(r/L)^{1-\zeta_{2,0}}$ 
with $\zeta_{2,0}=0.701$ (see Fig.\ref{gotoh_fig14}) \cite{Gotohetal02}. 
The nondimensional constant $\chi$ and the exponent $\alpha$ were found to 
be $0.027$ and $0.267$, respectively, by the least square fit 
in the range of $0.09\le r/L\le 0.3$. The latter value $0.267$ 
is slightly smaller than the value $1-\zeta_{2,0}=1-0.701=0.299$. 

From the above observation, we conclude that the conditional average 
$-\av{\delta p_{,x}|u,r}_N$ is well approximated by the quadratic function 
of $u$ with the curvature which decreases with the separation. 
The quadratic function of $u$ for $-\av{\delta p_{,x}|u,r}_N$ is consistent 
with the anti-shock action of the pressure. Locally the equation for $U$ 
may be described by the Burgers type equation with the pressure effect: 
\be
{DU\over Dt}\equiv \pt{U}{t}+U\pt{U}{r}
  =-\pt{\delta p}{X}+2\nu\pt{^2U}{r^2}+K(X,r,t),  \label{25}
\ee
\begin{figure}[t]
\refstepcounter{fgr}
\label{gotoh_fig\thefgr}
{\includegraphics[width=8.cm]{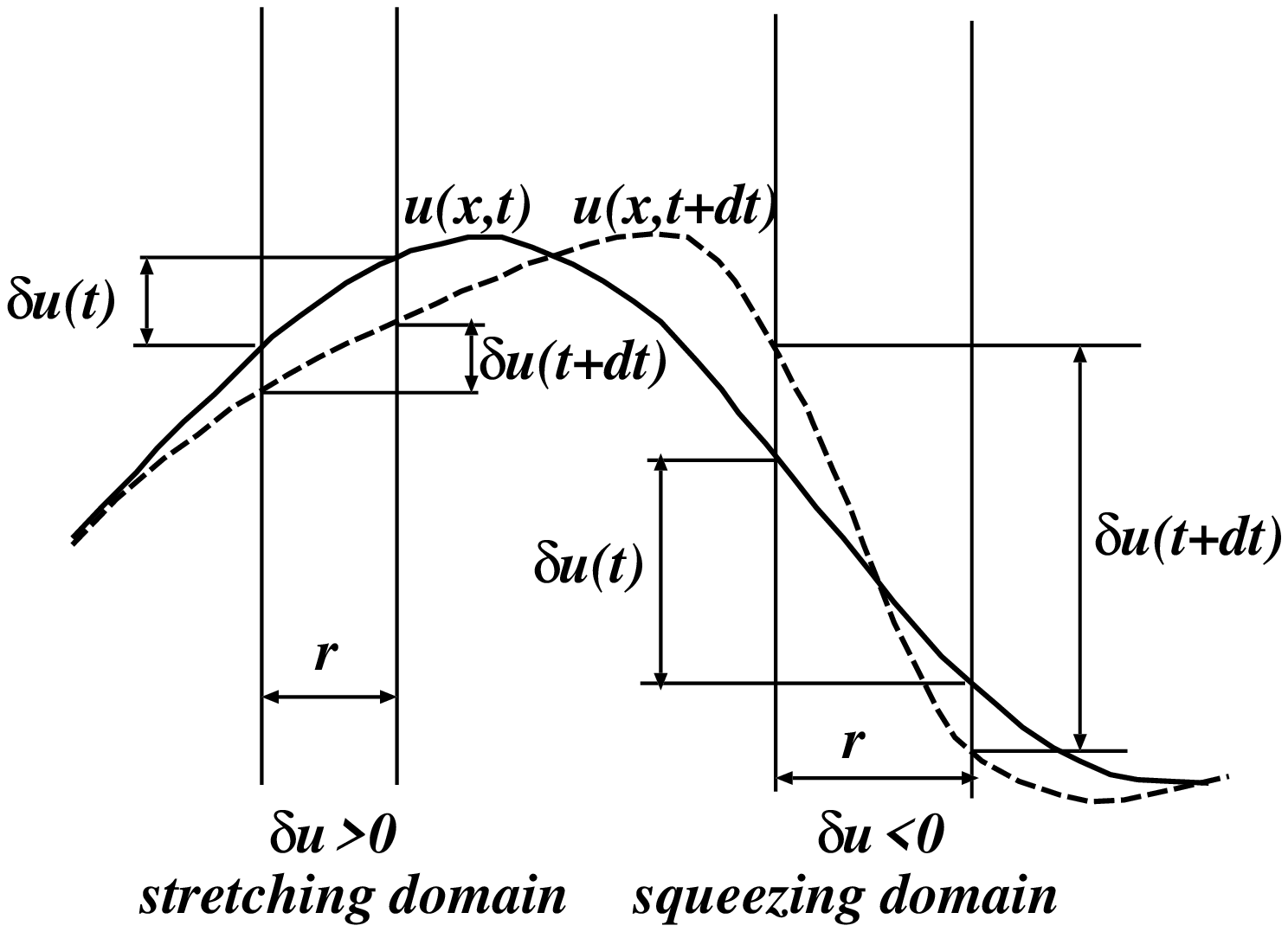}}

\vspace*{0.2cm}
\vbox{ 
\small Fig.\thefgr \ \  The role of the pressure term 
on the dynamics of the longitudinal velocity increment. 
The convection stretches the fluid portion with $\delta U>0$ 
($\xi=\partial U/\partial x$ decreases) and squeeze the 
fluid portion with $\delta U<0$ ($|\xi|$ increases). 
The modeled pressure gradient gives positive 
contribution $c_2U^2$ to $\partial U/\partial t$,  
that is, $\xi(>0)$ of the stretched portion is increased and 
$|\xi|(\xi<0)$ of the squeezed one is decreased. This means that 
the stretched and squeezed portions are pushed 
back left by the pressure gradient. }
\end{figure}%

\noindent
where $X$ is a $x$-coordinate moving with the average velocity at two points 
$\x_1$ and $\x_2$, and $K$ denotes the remaining 
terms arising from the Naveir-Stokes equation\cite{Fuka00}. 
If the pressure gradient term is modeled by the quadratic function 
of $U$ as inferred from the conditional average $-\av{\delta p_{,x}|U,r}$, 
all the terms except $K$ term are expressed in terms of $U$. 
This replacement is thought to be a kind of statistical projection 
of the pressure gradient onto $U$-space. 
The $K$ may be modeled as random force, but it is irrelevant 
in the following discussion.

To make discussion simpler, we take the leading term of 
$-\av{\delta p_{,x}|U,r}\approx a_2(r)U^2, a_2>0$. 
In the large negative $U$ region (squeezing domain), the convective term 
produces a shock as in Fig.\ref{gotoh_fig15}. 
The pressure gradient term makes a positive contribution 
$a_2U^2$ to $\partial U/\partial t$, which means that large (negative) 
amplitude $U(t+dt)$ is decreased. On the other hand, for large 
positive $U$ regions, those regions are stretched by 
the convective action so that the positive amplitude decreases\cite{rhk90,Gotohrhk93}. 
When the pressure gradient term is added, again this gives positive 
contributions to $\partial U/\partial t$,  
thus $U(t+dt)$ increases. Therefore the pressure gradient term acts in such 
a way that it suppresses the formation of singularity such as shocks by 
the convective motion, and this can be understood as 
the action which the curve $U(t+dt)$ is 
pushed back toward left by the pressure gradient in Fig.\ref{gotoh_fig15}.  
This implies that the pressure term is an indispensable 
ingredient for the dynamics of the longitudinal velocity increment. 

\section{Implication of the pressure gradient term}

When the closure \rf{19} is included in Eq.\rf{11} with $p=2n$ and $q=0$,  
the pressure-velocity correlation term becomes 
\ba
&& \hspace*{-0.8cm}
   -\av{(\delta p_{,x})U^{2n-2}}
     ={\hat{a}_2(r)\over r}\biggl(S_{2n,0}(r)
         -{S_{3,0}(r)\over S_{2,0}(r)}S_{2n-1,0}(r)
         \nonumber\\
&& \hskip 3cm
           -S_{2,0}(r)S_{2n-2,0}(r)\biggr).               \label{26}
\ea
Suppose that when the Ryenolds number and $n$ are very large, 
$S_{2n,0}(r)$ in Eq.\rf{26} is 
dominant in the inertial range.  
This requires $\zeta_{2n,0} \le \zeta_{2n-1,0}+1-\zeta_{2,0}$. 
On the other hand, the realizability condition requires that 
$\zeta_{p,0}$ is a nondecreasing function of $p$, meaning 
$\zeta_{2n-1,0}\le \zeta_{2n,0}$. Both constraints suggests that 
$0\le \zeta_{2n,0}-\zeta_{2n-1,0}\le 1-\zeta_{2,0}\approx 0.3$. In K41,  
$\zeta_{2n,0}-\zeta_{2n-1,0}=1/3$. From the DNS data, the above requirement 
is satisfied about at the order greater than four. 

Dominance of $S_{2n,0}$ over other two terms of Eq.\rf{26} means 
that Eq.\rf{11} becomes 
\be
  \dx{S_{2n,0}}{r}+{2\over r}S_{2n,0}
   ={2(2n-1)\over r}S_{2n-2,2}+(2n-1){\hat{a}_2(r)\over r}S_{2n,0},  \label{27}
\ee
asymptotically in the inertial range. 
Suppose that $\hat{a}_2(r)\propto r^{\gamma}$ in the inertial range. 
When $\gamma<0$, since the two terms of the right hand side of Eq.\rf{27} 
are positive and the last term becomes dominant, the both side can not balance. 
This means that the case of $\gamma<0$ can not occur. 
When $\gamma=0$, both sides of Eq.\rf{27} balances, 
thus we have $\zeta_{2n,0}=\zeta_{2n-2,2}$. 
If $\gamma>0$, the last term of the right hand side of Eq.\rf{27} becomes 
less important so that the left hand side balances with the first term 
of the right hand side and $\zeta_{2n,0}=\zeta_{2n-2,2}$. 
The conclusion of the above arguments is $\zeta_{2n,0}=\zeta_{2n-2,2}$ for large 
$n$, but this equality is totally dependent on the assumption of the quadratic function 
in $U$ for the conditional average (see also the discussion by Hill\cite{Hill01}).  

Although the DNS data shows that $\hat{a}_2(r)=\chi C_0^{-1}(2F_{\nabla p}/3)^{1/2}$ 
is approximately constant, there is small but finite difference between the exponent 
$\alpha=0.267$ and $1-\zeta_{2,0}=0.299$. If we regard these two values 
correct, we have $\gamma=0.299-0.267=0.032>0$. However, we think that our data 
is not accurate enough for definite judgement, and the preceding 
argument about the role of the pressure term suggests that $\gamma=0$ 
is very plausible.  

It is important and interesting to seek physical explanation for possible 
$r$-dependence of $\hat{a}_2(r)$, if it is the case. 
To the authors' best knowledge, 
no theory has been developed to compute $\hat{a}_2(r)$ including a nondimensional 
constant.  We present one possible argument in the following. 

The pressure is given by the spatial integral of the source term 
which is quadratic in the velocity gradients, so that the pressure gradient 
is given by 
\ba
&&   \nabla p=-\int_V \nabla_{\x}G(\x-\y)B(\y) d\y, 
    \label{28} \\
&&   G(\x-\y)=-{1\over 4\pi} {1\over |\x-\y|},            \label{29}
\ea
for the periodic boundary condition in the present study, 
where $B=\nabla\u:\nabla\u$. 
The integral is roughly evaluated as 
the product of $B$, $\nabla G$ and an effective volume $V_{effect}$\cite{Gr99}.  
The derivative of the  Poisson kernel $\nabla G$ does not yield non-integer power, 
and the source term $B$ would contribute to $S_{2n,0}$. 
The effective volume is the one over which the source term has 
a dominant contribution to the integral, and would change its shape  
and size depending on the amplitude and its spatial coherency of $B$. 
This effective volume is characterized by a geometrical non-integer dimension.   
Therefore it is plausible that the effective volume $V_{effect}$ may 
induce a non-integer power law dependence of $\hat{a}_2(r)$ on the scale $r$. 

\section{Summary}

We have examined the dynamical relations for the various kinds 
of the velocity structure functions in terms of the DNS data. 
It was argued that the pressure term is a key to understand the 
scaling of the velocity structure functions. 
The contribution of the pressure gradient is important to 
the dynamics of the longitudinal structure functions, 
while in the case of the transverse or mixed ones the pressure 
gradient may or may not be important. 

To assess the statistical role of the pressure gradient term 
in the dynamics of the longitudinal velocity increment, we have 
computed the conditional average of the pressure gradient increment 
with a given value of $U$ and $r$. It was found that the quadratic 
function in $U$ is a good approximation to 
the conditional average. 
By applying two constraints to the conditional average and using 
Kolmogorov's 4/5 law, we have theoretically determined 
two coefficients of the quadratic function but there remains one 
undetermined coefficients. 

The role of the pressure gradient was argued in terms of the 
equation for $U$ which is similar to the Burgers' equation 
but with the term which models the pressure gradient. 
The pressure gradient acts to resist to stretching and squeezing 
of the fluid element, leading to shock smearing, which 
in turn means that the scaling exponent of $S_{2n,0}$ is not  
constant in $n$, unlike the Burgers case, and is a slowly increasing 
function of the order. 
This means that the pressure is an intermittency killer
(Kraichnan 1991)\cite{rhk91}. 

The quadratic function form for the conditional average 
of the pressure gradient increment leads to the closure of the moment 
equation for $S_{2n,0}$. If higher order polynomial form is 
used, the closure fails. However, if a thin vortex tube is 
the most singular object under the Navier-Stokes dynamics, 
the argument using the Bernoulli theorem nearby the tube described 
in Sec.4 suggests that there are no higher order terms than the second 
order. The bottom line is $\zeta_{2n,0}=\zeta_{2n-2,2}$.  
It was argued that the effective volume 
is an important factor for the dependence of $\hat{a}_2(r)$ on $r$. 
Further theoretical study is necessary. 
The conditional average of the pressure gradient also leads to 
the closure of the Liouville equation for the PDF of $U$. 
This analysis is now underway and will be reported elsewhere. 

\vskip 0.5cm

We thank to J. Davoudi, R. J. Hill, R. H. Kraichnan, S. Kurien, 
K. Sreenivasan, and V. Yakhot for their useful comments. 
The authors are grateful to Dr. Fukayama and Mr. Kajita for 
their assistance of the computation. 
The authors wish to thank the Nagoya University 
Computation Center, the Advanced Computing Center at RIKEN  and 
the Computer Center of the National Fusion 
Science of Japan for providing the computational resources. 
This work was supported by a Grant-in-Aid for Scientific Research 
(C-2 12640118) from  Japan Society for the Promotion of Science.


\end{document}